\documentstyle[poliface]{article}

\renewcommand{\textwidth}{16.5cm}

\newcommand{\Ncollisions}{{\cal N}_{\rm coll}}

\newcommand{\Ne}{N_e}

\newcommand{\re}{r_{\rm e}}
\newcommand{\rsmallb}{r_{\rm b}}

\newcommand{\Rt}{{\cal R}_t}
\newcommand{\Rtdot}{\dot{\Rt}}

\newcommand{\Sfour}{S^{(4)}}
\newcommand{\Sone}{S^{(1)}}

\newcommand{\Sdil}{S$^{\rm dil}$}
\newcommand{\Sconc}{S$^{\rm conc}$}

\newcommand{\ta}{t_a}
\newcommand{\th}{t_h}
\newcommand{\tstartwo}{t_{2}^{*}}
\newcommand{\tstarmany}{t_{m}^{*}}

\newcommand{\tstarmanyR}{t_{m,R}^{*}}
\newcommand{\tl}{t_l}

\newcommand{\te}{t_{\rm e}}
\newcommand{\tb}{t_{\rm b}}

\newcommand{\nainf}{n_{A}^{\infty}}
\newcommand{\nbinf}{n_{B}^{\infty}}

\newcommand{\rhoabs}{\rho _{AB}^{s}}

\newcommand{\tsat}{t_{\rm sat}}

\newcommand{\Rcrit}{{\cal R}_{\rm crit}}

\newcounter{fignumber}

\begin{document}



\renewcommand{\thepage}{}

\titleben{\LARGE \bf Reactions at Polymer Interfaces: Transitions from Chemical
to Diffusion-Control and Mixed Order Kinetics}

\author{\Large 
BEN O'SHAUGHNESSY\ $^{1}$ \ and \ DIMITRIOS VAVYLONIS\ $^2$ \\ 
}

\maketitle

\ \\ \newline
{\large $^1$ Department of Chemical Engineering, Columbia University,
500 West 120th St., New York, NY 10027, USA} \\
\ \\
{\large $^2$ Department of Physics, Columbia University, 538 West
120th St., New York, NY 10027, USA} \\
\ \\ \ \\
\ \\ \ \\ \ \\

\pagebreak


\pagenumbering{arabic}

\large

\section*{ABSTRACT}

We study reactions between end-functionalized chains at a
polymer-polymer interface.  For small chemical reactivities (the
typical case) the number of diblocks formed, $\Rt$, obeys 2nd order
chemically controlled kinetics, $\Rt \twid t$, until interfacial
saturation.  For high reactivities (e.g. radicals) a transition occurs
at short times to 2nd order diffusion-controlled kinetics, with $\Rt
\twid t/\ln t$ for unentangled chains while $t/\ln t$ and $t^{1/2}$
regimes occur for entangled chains.  Long time kinetics are 1st order
and controlled by diffusion of the more dilute species to the
interface: $\Rt\twid t^{1/4}$ for unentangled cases, while $\Rt\twid
t^{1/4}$ and $t^{1/8}$ regimes arise for entangled systems.  The final
1st order regime is governed by center of gravity diffusion, $\Rt\twid
t^{1/2}$.

\viv
\vii

PACS numbers:
  \begin{benlistdefault}
    \item [ 82.35.+t ]
(Polymer reactions and Polymerization)
    \item [05.40.+j]
 (Fluctuation Phenomena, random processes, and Brownian
Motion)
     \item [68.10.-m]
 (Fluid Surfaces and Fluid-Fluid Interfaces)
\end{benlistdefault}

\pagebreak


In many technological applications, interfaces in polymeric
multicomponent materials are reinforced through chemical bonding
\citeben{sperling:book}.  Techniques of reinforcing interfaces
separating immiscible polymer species involve attaching functional
groups to a certain fraction of the bulk chains
\citeben{becktan:morph_cont_reinforce_blends,kramer:strength_iface_faraday}.  Under melt
conditions, reactions produce copolymers (see fig.
\ref{pol_iface_letter}) which serve as interfacial bridges, enhancing
interfacial fracture toughness and yield stress after cooling
\citeben{gersappe:bind_blends_copolym}.  In commercial applications 
(``reactive processing'') the two melts are
simultaneously mechanically mixed.  In addition to their direct
reinforcing effect, the copolymer products also help to produce a
finer blend morphology by reducing interfacial tension and preventing
the coalescence of minority phase droplets during mixing
\citeben{becktan:morph_cont_reinforce_blends}.

Due to the importance of such applications, recent theoretical
and numerical
research efforts have focused on understanding the reaction kinetics
at polymer-polymer interfaces at a fundamental level
\citeben{ben:reactiface_uday_all_europhys,fredrickson:reactiface_prl,%
fredricksonmilner:reactiface_timedept,muller:reactiface_montecarlo},
in the simplest case of end-functionalized chains (see fig.
\ref{pol_iface_letter}).  These studies have addressed only a small
fraction of the available parameter space.  They emphasized the limit
of ``infinitely'' reactive groups, \ie local chemical reactivities $Q$
of order $1/\ta$ where $\ta$ is the monomer relaxation time.  In fact,
such $Q$ values are realized only for exotic species such as radicals
(or electronically excited groups as in model photophysical systems)
whilst chemical reactivities employed in most reactive blending
experiments are extremely small.  A typical example is the reaction
between carboxylic acid and epoxide ring groups for which $Q \ta
\approx 10^{-11}$
\citeben{guegan:reactiface_kinetics}.
In addition, with the exception of ref.
\citenum{fredricksonmilner:reactiface_timedept}, previous studies have
been confined to reactive chains very dilute in an unreactive melt
background.

In this letter we develop a theory for the kinetics of reactions
between end-functionalized chains at a stationary interface separating
immiscible melts A and B (see fig.
\ref{pol_iface_letter}).  Our approach is quite general in terms of chemical
reactivity, molecular weight, and number densities of reactive ends in
each bulk, $\nainf$ and $\nbinf$.  The convention throughout is that A
reactants are more dilute, $\nainf\le\nbinf$.  We will construct
``phase diagrams,'' representing various classes of reaction kinetics,
in the density-reactivity plane for both unentangled and entangled
polymer melts.

The principal conclusions of this study are as follows: (1) For the
usual case of small reactivities, $Q \ta \lsim 10^{-6}$, the behavior
is as follows.  Chemically controlled 2nd order kinetics apply at
short times, with number of reactions per unit area, $\Rt$, increasing
as $\Rt \twid t\, \nainf \nbinf$.  If $\nainf \approx
\nbinf$, these kinetics persist until crowding of the interface by A-B diblocks
strongly suppresses further reactions.  However, for $\nainf$
sufficiently smaller than $\nbinf$, then before the interface gets
saturated a crossover occurs to 1st order diffusion-controlled (DC)
kinetics with $\Rt \approx x(t)\, \nainf$ where $x(t)$ is the rms
monomer displacement in time $t$.  (2) For the case of highly reactive
groups, $Q \ta\approx 1$, irrespective of density a crossover occurs
from 2nd to 1st order kinetics at long times.  In addition, a new 2nd
order DC regime appears during which $\Rt \approx x^4(t)\, \nainf
\nbinf$.

Our aim is the reaction rate, ${d \over dt} \Rt$.  To begin, we
observe that this requires calculation of the 2-body correlation
function $\rhoabs$, proportional to the number of A-B reactive groups
in contact at the interface: ${d \over dt} \Rt = Q h a^3
\rhoabs(t)$.  Here $h$ is the width of the interface and $a$ the monomer
size.  Unfortunately, the exact dynamical equation for $\rhoabs$
involves 3-body correlations.  It
is in fact impossible to write a closed exact equation for $\rhoabs$
since correlation functions of all orders are coupled in an infinite
hierarchy of dynamical equations.  This complication which arises in
all many-body reacting systems is typically resolved by approximating
3-body correlations in terms of lower order correlation functions
\citeben{kotominkuzovkov:book_short}.  However, in ref.
\citenum{ben:reactiface_fund_note} (where a general theory for interfacial
reaction kinetics was developed) after postulating much less
restrictive bounds on the magnitude of 3-body correlation functions we
obtained the following closed self-consistent equation for $\rhoabs$:
                                                \begin{eq}{door}
\rhoabs(t) \approx \nainf \nbinf - 
                \lambda \int_0^t dt'\, \Sfour_{t-t'}\, \rhoabs(t') 
              - \lambda \nbinf \int_0^t dt'\, \Sone_{t-t'}\, \rhoabs(t') 
\comma
\gap 
\lambda \equiv Q h a^3
\comma
                                                                \end{eq}
where $\Sfour_t \approx 1/x^{4}(t)$ is the probability density
\citeben{ben:reactiface_uday_all_europhys} that in the absence of
reactions an A-B pair is in contact at the interface at $t$, given its
members were in contact at the interface initially.  $\Sone_t \approx
1/x(t)$ is the probability density an A or B group, initially at the
interface, returns to it after time $t$.
(Eq. \eqref{door} is closed in terms of the degrees of freedom
describing the reactive end-groups only.  To achieve this, it has
implicitly been assumed that reactions do not modify equilibrium
internal polymer chain configurations
\citeben{ben:intersemi_all_europhys}.
There is considerable evidence from renormalization group analyses
\citeben{ben:interdil_all_europhys} that this approximation
correctly captures all scaling dependencies, though prefactors are
unreliable.)  Eq. \eqref{door} is a closed system in terms of
$\rhoabs$; though spatial degrees of freedom in the bulk have been
integrated out, the behavior of $\rhoabs$ reflects bulk correlation
functions.  For example, the signature in eq. \eqref{door} of a
reaction-induced density ``hole'' near the interface is domination of
the $\nainf\nbinf$ and $\Sone$ terms (these 2 terms then balance one
another).  On the other hand, a depletion hole in the bulk {\em
2-body} correlation function corresponds to a balance of the
$\nainf\nbinf$ and $\Sfour$ terms \citeben{ben:reactiface_fund_note}.  In
the following, instead of directly solving eq. \eqref{door} for
$\rhoabs$, we will rather motivate the solutions using physical
arguments.  The reader may then verify the results by direct
substitution into eq. \eqref{door}.

Consider the reaction rate at short times.  Initially, reactive groups
are distributed as in equilibrium, $\rhoabs \approx \nainf \nbinf$.
Hence initial kinetics are 2nd order, $\Rt \approx \lambda t\, \nainf
\nbinf$.  However, reactions may perturb equilibrium correlations.
Consider times shorter than $\tl$, the diffusion time corresponding to
a distance of order the typical separation between the more dense B
reactive groups, $x(\tl) = (\nbinf)^{-1/3}$.  For $t<\tl$ reactions
are due to the few isolated A-B pairs which happened to be initially
within diffusive range, \ie within $x(t)$ of one another near the
interface (see fig. \ref{pol_iface_letter}).  After time $t$, the
number of times such a pair has collided, $\Ncollisions$, is of order
the number of encounters the A makes with the interface, $(t/\ta)
(h/x(t))$, multiplied by the probability it meets the B, $a^3/x^3(t)$,
during each of these encounters.  Hence the total reaction
probability, $Q \ta
\Ncollisions\approx \lambda t /x^4(t)$ is increasing with time
provided the dynamical exponent $z$ characterizing monomer diffusive
dynamics, $x(t) \twid t^{1/z}$, is greater than or equal to 4 (the
case $z=4$ is marginal, but it can be shown to be similar to $z>4$).
Now for times shorter than $\tau$, where $\tau$ is the longest polymer
relaxation time, values of $z$ are either 4 or 8
\citeben{doiedwards:book}.  This implies that the pair reaction
probability is increasing with time and hence there exists a timescale
$\tstartwo$, obeying $x^4(\tstartwo)/ \tstartwo = \lambda$, after
which all such A-B pairs will have reacted.  Thus for $t>\tstartwo$
non-equilibrium correlations develop ($\rhoabs$ is no longer equal to
$\nainf \nbinf$) and a depletion hole grows in the 2-body correlation
function.  $\Rt$ is then proportional to the number of pairs per unit
area which diffusion could have brought together by time $t$, $\Rt
\approx x^4(t) \nainf \nbinf$.  These are 2nd order diffusion
controlled kinetics.

For how long do these DC kinetics persist?  For $t>\tl$, since at
least one B lies within the exploration volume of any A within $x(t)$
of the interface, thus any such A must have reacted.  For $t>\tl$, a
depletion hole thus grows in the A density field and the reaction rate
becomes controlled by the diffusion of A species to the interface, $\Rt
\approx x(t) \nainf$.  These are 1st order DC kinetics.  In summary,
                                                \begin{eq}{sconc}
\Rt \, \approx\, 
\lambda\, t\, \nainf \nbinf\ 
\stackrel{\tstartwo}{\ggt} 
\ x^4(t)\, \nainf \nbinf\ 
\stackrel{\tl}{\ggt} 
\ x(t)\, \nainf 
\gap 
(\tstartwo < \tl < \tau,\  \ \mbox{\Sconc}) 
\period
                                                                \end{eq}
Notice that while the more dilute side controls the long time
reaction rate, it is the more dense which determines the crossover
time $\tl$.  Later, we will return to eq. \eqref{sconc} with explicit
expressions for $x(t)$.

Certain assumptions have been made in deriving eq. \eqref{sconc}.  It
was implicit that $\tstartwo<\tl<\tau$ since for $t>\tau$ chain center
of gravity Fickian diffusion takes over and $x(t) \twid t^{1/2}$, \ie
$z=2$ and $\Ncollisions$ no longer increases with time; hence our
previous arguments are invalid unless both $\tstartwo$ and $\tl$ occur
before $\tau$.  Our arguments also assumed that $\tstartwo<\tl$; were
this not true, then an A group could have encountered many B's and yet
remain unreacted.   These constraints limit the
validity of eq. \eqref{sconc} to sufficiently high $Q$ and $\nbinf$
only.  In the $Q$-$\nbinf$ plane, these conditions are satisfied
in the region marked \Sconc\ (for strongly reactive, concentrated) in figs.
\ref{iface_rouse_letter} and \ref{iface_rept_letter}.

Let us see first how the kinetics of eq. \eqref{sconc} are modified
when $Q$ and $\nbinf$ have values such that $\tstartwo>\tl$.  Consider
an A group initially within $x(t)$ of the interface.  By time $t$ it
collides with the interface of order $(t/\ta) (h/x(t))$ times.  Now
since $\tstartwo>\tl$, its reaction probability by $\tl$ must be
small; this means each such group collides with {\em many} B's before
it reacts.  Now each encounter with the interface produces reaction
with probability $\approx \nbinf a^3 Q\ta$.  The net reaction
probability $\lambda \nbinf t/x(t) \twid t^{1-1/z}$ is thus increasing
with time and becomes of order unity at $\tstarmany$, where
$x(\tstarmany)/ \tstarmany = \lambda \nbinf$.  It follows that for
$t>\tstarmany$, all A groups within diffusive range of the interface
will have reacted and a crossover occurs to 1st order DC kinetics:
                                                \begin{eq}{weak}
\Rt \, \approx\, 
\lambda\, t\, \nainf \nbinf\ 
\stackrel{\tstarmany}{\ggt} 
\ x(t)\, \nainf 
\gap 
(\tstartwo > \tl \ \mbox{or} \ \tstartwo>\tau, \ \ \mbox{W}) 
                                                                \end{eq}
Notice the absence of 2nd order DC kinetics.  In fact, one can show
eq. \eqref{weak} is also valid if $\tstartwo>\tau$.  This is because
reaction probability for a single pair always stops increasing after
$\tau$.  In the $Q$-$\nbinf$ plane of figs.
\ref{iface_rouse_letter} and \ref{iface_rept_letter}, the conditions
under which eq. \eqref{weak} is valid correspond to region W, for
``weakly reactive'' (small $Q$).

The remaining region of the $Q$-$\nbinf$ plane is that labeled \Sdil\
(``strongly reactive, dilute'') in figs. \ref{iface_rouse_letter} and
\ref{iface_rept_letter}, corresponding to $\tstartwo<\tau<\tl$.  In
this region all arguments leading to eq.
\eqref{sconc} are still valid for $t<\tau$.  But for $t>\tau$, $z=2$
and reaction is no longer guaranteed when 2 groups are within
diffusive range.  Thus the hole in the 2-body correlation function
stops growing and the 2nd order DC kinetics cease.  However, since
$\tstartwo<\tau$, reaction is certain when two coils overlap at the
interface.  Since by $\tau$ there is no density depletion at the
interface, thus for $t>\tau$, the reaction rate is proportional to the
equilibrium number of overlapping coils per unit area at the
interface: $\Rt \approx (R^4/\tau) t\, \nainf \nbinf$, where $R$ is
the polymer coil radius.  Repeating the argument leading to the
calculation of $\tstarmany$, but replacing $a$ and $h$ by $R$ and $Q$
by $1/\tau$, leads to the conclusion that the reaction rate becomes
limited by delivery of A species to the interface at a timescale
$\tstarmanyR = \tau (R^3 \nbinf)^{-2}$.  In summary:
                                                \begin{eq}{sdil}
\Rt \, \approx\, 
\lambda\, t\, \nainf \nbinf\ 
\stackrel{\tstartwo}{\ggt} 
\ x^4(t)\, \nainf \nbinf\ 
\stackrel{\tau}{\ggt} 
(R^4/\tau) t\, \nainf \nbinf\
\stackrel{\tstarmanyR}{\ggt} 
\ x(t)\, \nainf 
\gap 
(\tstartwo < \tau < \tl,\  \ \mbox{\Sdil}) 
                                                                \end{eq}
This completes the derivation of our general results, eqs.
\eqref{sconc}, \eqref{weak} and \eqref{sdil}. 
Let us now apply these to unentangled and entangled melts,
respectively.

{\em 1. Unentangled chains.} For shorter chain lengths $N$, Rouse
dynamics apply \citeben{doiedwards:book}, $x(t) \twid t^{1/4}$ for
$t<\tau \approx N^2 \ta$. Ordinary Fickian diffusion, $x(t) \twid
t^{1/2}$, onsets for $t>\tau$.  Following the steps described above,
the phase diagram of fig. \ref{iface_rouse_letter} is constructed.
Explicit formulae for timescales are obtained after inserting the
appropriate $x(t)$ forms.  The only complication is that the short
time $z=4$ regime is marginal ($\Ncollisions$ increases only
logarithmically with time).  This leads to logarithmic corrections
during the 2nd order $x^4(t)$ DC kinetics:
                                                \begin{eq}{log}
\Rt \approx a^4 t/[\ta \ln (t/\th)] \nainf \nbinf \ggap
(\mbox{2nd order DC})
\comma
                                                                \end{eq}
where $\th$ is the diffusion time corresponding to $h$.  Similarly, the
$(R^4/\tau) t\, \nainf \nbinf$ regime is modified to $\Rt \approx R^4 t/[\tau
\ln(\tau/\th)] \nainf \nbinf$.  These logarithmic corrections also 
modify the timescales $\tstartwo, \tl$ and $\tstarmanyR$.  From the
results presented here, they are easily determined by demanding
continuity of reaction rates.

As an example, consider a point in fig. \ref{iface_rouse_letter}
belonging to region \Sconc.  The kinetic sequence is:
                                                \begin{eq}{apple}
\Rt \, \twid \, 
t\, \nainf \nbinf\ 
\stackrel{\tstartwo}{\ggt} 
\ (t/\ln t)\, \nainf \nbinf\ 
\stackrel{\tl}{\ggt} 
\ t^{1/4} \nainf
\stackrel{\tau}{\ggt} 
\ t^{1/2} \nainf
\period
                                                                \end{eq}

{\em 2. Entangled chains.} For $N$ above the threshold $\Ne$,
entanglements onset.  According to the ``tube'' or ``reptation'' model
\citeben{doiedwards:book} Rouse dynamics apply for 
$t<\te \equiv \Ne^2\ta$ until $x(t)$ reaches the tube diameter
$\re\equiv \Ne^{1/2} a$.  The ``breathing modes'' $t^{1/8}$ regime
follows, lasting until $\tb \equiv N^2\ta$, while $x(t)\twid t^{1/4}$
for $\tb<t<\tau$ where $\tau \equiv (N^3/\Ne)\ta$.  Diffusion is
Fickian for $t>\tau$.

Similarly to the unentangled case one constructs the phase diagram of
fig. \ref{iface_rept_letter}.  The 3 basic regions \Sconc, \Sdil\ and W
now develop fine structure since each characteristic timescale may
belong to different $x(t)$ regimes.  (For example, for point X of fig.
\ref{iface_rept_letter} one has $\te<\tstartwo<\tb$, and
$\tb<\tl<\tau$.)  Again, logarithmic corrections to the 2nd order DC
kinetics arise during the two short time $t^{1/4}$ regimes, with $\Rt
\approx a^4 \ t/ [\ta \ln(t/\th)] \nainf \nbinf$ during the first, and
$\Rt \approx \rsmallb^4\, t/[\tb \ln( t/\tb)] \nainf \nbinf$ during
the second (where $\rsmallb \equiv N^{1/4} \Ne^{1/2} a$).

As an example consider point X marked in fig. \ref{iface_rept_letter},
belonging to the \Sconc\ regime.  Then eq. \eqref{sconc} implies:
                                                \begin{eq}{orange}
\Rt \, \twid \, 
t\, \nainf \nbinf\ 
\stackrel{\tstartwo}{\ggt} 
\ t^{1/2} \nainf \nbinf
\stackrel{\tb}{\ggt}
\ (t/\ln t)\, \nainf \nbinf\ 
\stackrel{\tl}{\ggt} 
\ t^{1/4} \nainf
\stackrel{\tau}{\ggt} 
\ t^{1/2} \nainf
\period
                                                                \end{eq}

Finally, let us consider effects of interfacial saturation by
products.  When $\Rt$ reaches the value $\Rcrit \equiv N^{-1/2}
a^{-2}$, the diblocks start to feel one another laterally and
therefore begin to stretch
\citeben{ben:reactiface_uday_all_europhys,semenov:block_copol_microphase}.
Now during the very late stages, when $\Rt\gg \Rcrit$, incoming chains
must overcome a very large free energetic barrier in order to react
with the surface.  It was shown in ref.
\citenum{ben:reactiface_uday_all_europhys} that this leads to a
reaction rate decreasing exponentially with surface coverage, $\Rtdot \twid
e^{-9(\Rt/\Rcrit)^2}$, which in turn implies that $\Rt$ increases
logarithmically slowly in time, as demonstrated in ref.
\citenum{fredricksonmilner:reactiface_timedept}.  (For
coverages greater than but of order $\Rcrit$ one expects $\Rtdot \twid
f(\Rt/\Rcrit)$ where $f(u\gg 1)= e^{-9 u^2}$.  The detailed form of
the cross-over function $f$ describes this intermediate regime.  $f$
is in principle available from eq. \eqref{door}.)

Thus, for very large $N$, in effect reactions cease at coverages large
compared to $\Rcrit$ (here we neglect destabilization effects which
may increase interfacial area).  Equating $\Rt$ to $\Rcrit$, we thus
find that for all points well above a certain line (labeled
$\tstarmany=\tsat$) in figs.  \ref{iface_rouse_letter} and
\ref{iface_rept_letter}, saturation occurs during the final 1st order
DC kinetics regime at a time $\tsat= \tau/(N \nainf a^3)^2$.  Below
that line, $\tsat = 1/(Q \nainf \nbinf h a^4 R)$ is smaller than
$\tstarmany$ and saturation occurs before the crossover to the 1st
order DC regime.

We conclude with a brief discussion of our main findings.  Since
ordinary bimolecular rate constants correspond to $Q \ta \lsim
10^{-6}$, experimental systems typically lie below the $\tsat =
\tstarmany$ line of figs. \ref{iface_rouse_letter} and
\ref{iface_rept_letter}.  Then, apparently consistent with
experiment \citeben{scottmacosko:model_react_processing}, linear 2nd
order kinetics, $\Rt \twid t$, apply until surface crowding by
products at $\tsat$.  As a specific example, if all chains were
functionalized, $N = 300$ and $Q \ta \approx 10^{-6}$, then $\tsat$
would be of order seconds.  Typically groups are even less reactive
and $\tsat$ much longer.  Note that if $\nainf\ll\nbinf$ then even for
small $Q\ta$ a transition occurs to 1st order DC kinetics.  However
even in this case these kinetics may not be observable since
$\tstarmany$ is likely to exceed reactive blending experimental
timescales (of order minutes).  We conclude that typical reactive
blending experiments involving end-functionalized polymers are not in
the DC regime.  This would seem to be consistent with recent
experiment \citeben{orr:reactive_selfassembly}.

Of considerable potential interest are experiments involving highly
reactive radical species or strongly physically associating polymers,
for which $Q \ta \approx 1$.  For these, a range of 2nd and 1st order
DC regimes are predicted.  Measuring these various kinetic regimes
with laser-induced radicals thus offers an interesting probe of
fundamentals of interfacial polymer dynamics.

\centerline{***}

This work was supported by the National Science Foundation, grant no.
DMR-9403566.  We thank Uday Sawhney for stimulating discussions.


\pagebreak


\pagebreak


                     \begin{thefigures}{99}


\figitem{pol_iface_letter}

End-functionalized chains reacting at an interface separating
immiscible melts A, B containing functional groups, of local
reactivity $Q$, whose initial densities are $\nainf, \nbinf$.
For short times, reactions are confined to those groups whose
exploration volumes of size $x(t)$ overlap at the interface.  The
number of such pairs per unit area is $x^4(t) \nainf \nbinf$.  


\figitem{iface_rouse_letter}

Unentangled melts, reaction kinetics ``phase diagram'' in $Q-\nbinf$
plane.  Axes are logarithmic and units chosen such that $\ta=a=1$.
Maximum possible density is $\nbinf a^3=1/N$ (all chains
functionalized).  Kinetics are qualitatively distinct in the 3 regions
\Sconc, \Sdil and W (defined by solid lines).
The \Sconc/W boundary defines a critical density-dependent reactivity,
$Q=Q^*$.  Region W is divided into two sub-regions by the
$\tstarmany=\tau$ line, above (below) which $\tstarmany<\tau$
($\tstarmany>\tau$).  The line $\tsat = \tstarmany$ is shown for the
special case of constant $\nainf/\nbinf$ (the convention is $\nainf
\leq \nbinf)$.


\figitem{iface_rept_letter}

As fig. \ref{iface_rouse_letter}, but for entangled melts.  Regions
\Sconc, \Sdil and W now develop fine structure.  In a given sub-region, 
relevant timescales occur within a given reptation diffusion regime
thus defining a unique sequence of kinetic regimes.


                     \end{thefigures}

\pagebreak


%
%
%
%
%

\begin{figure}[t]

\epsfxsize=\textwidth \epsffile{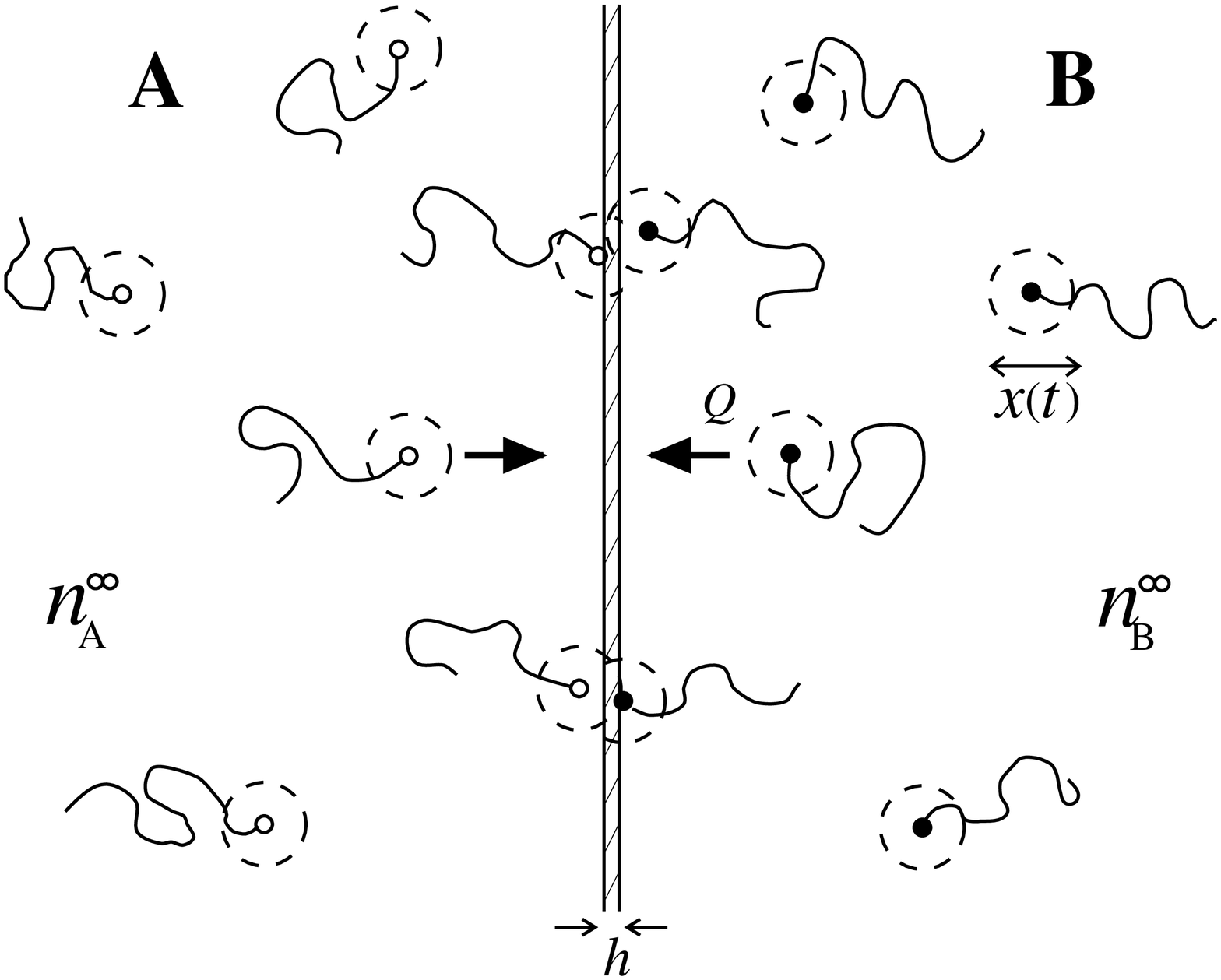}

\end{figure}

\mbox{\ }

\vfill

\addtocounter{fignumber}{1}
\mbox{\ } \hfill {\huge Fig.\@ \thefignumber} 

\pagebreak
\begin{figure}[t]

\epsfxsize=\textwidth \epsffile{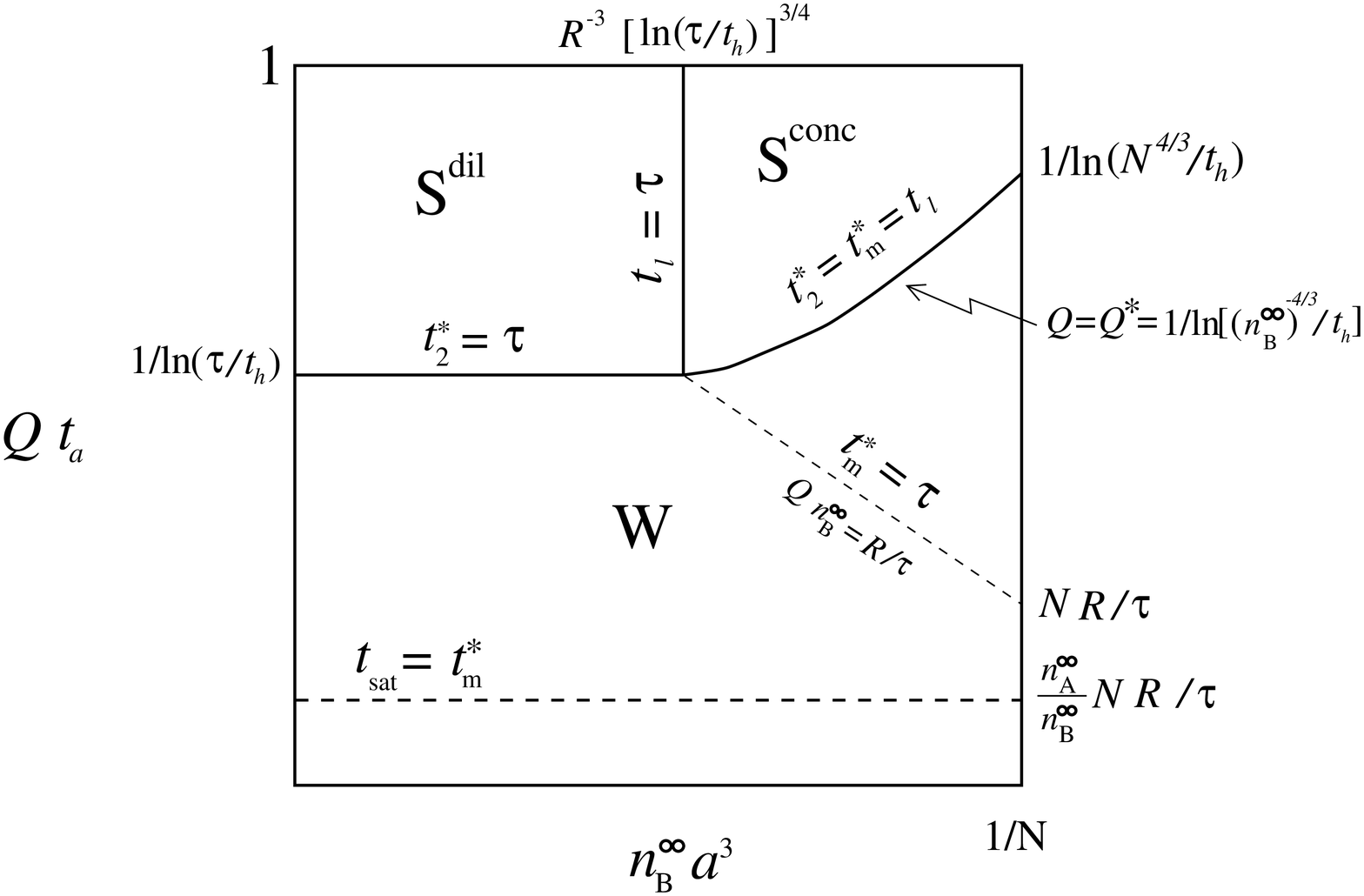}

\end{figure}

\mbox{\ }

\vfill

\addtocounter{fignumber}{1}
\mbox{\ } \hfill {\huge Fig.\@ \thefignumber} 

\pagebreak
\begin{figure}[t]

\epsfxsize=\textwidth \epsffile{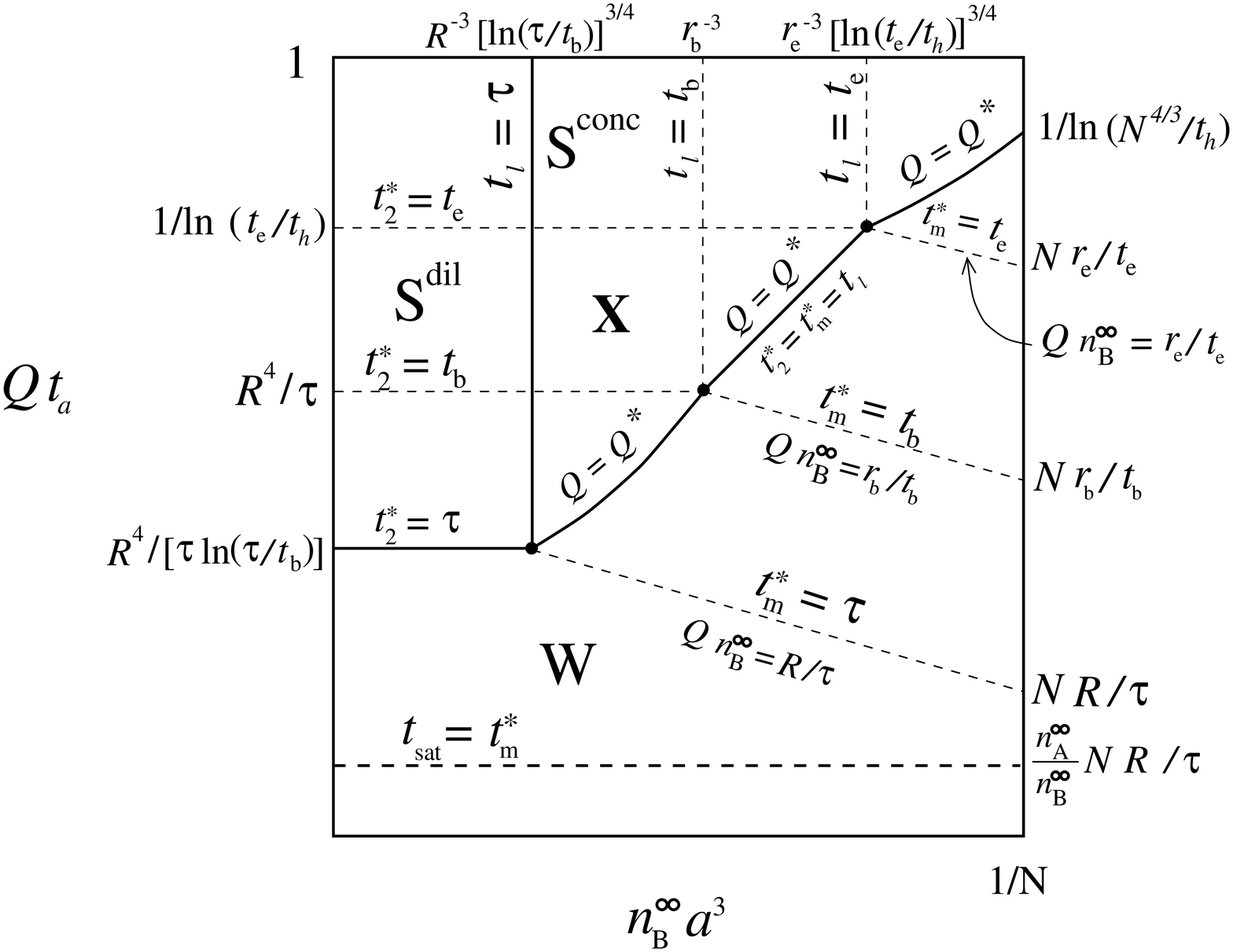}

\end{figure}

\mbox{\ }

\vfill

\addtocounter{fignumber}{1}
\mbox{\ } \hfill {\huge Fig.\@ \thefignumber} 

\pagebreak

\end{document}